\documentstyle[12pt]{article}
\begin{document}

\def\a{\alpha}
\def\b{\beta}
\def\ch{\chi}
\def\d{\delta}
\def\e{\epsilon}
\def\f{\phi}
\def\g{\gamma}
\def\h{\eta}
\def\i{\iota}
\def\j{\psi}
\def\k{\kappa}
\def\l{\lambda}
\def\m{\mu}
\def\n{\nu}
\def\o{\omega}
\def\p{\pi}
\def\q{\theta}
\def\r{\rho}
\def\s{\sigma}
\def\t{\tau}
\def\u{\upsilon}
\def\x{\xi}
\def\z{\zeta}
\def\D{\Delta}
\def\F{\Phi}
\def\G{\Gamma}
\def\J{\Psi}
\def\L{\Lambda}
\def\O{\Omega}
\def\P{\Pi}
\def\S{\Sigma}
\def\U{\Upsilon}
\def\X{\Xi}
\def\T{\Theta}

\def\xt{\bar{x}}
\def\yt{\bar{y}}
\def\wt{\bar{w}}
\def\zt{\bar{z}}
\def\vt{\bar{v}}
 \def\pp {\partial }
\def\pb {\bar{\partial }}
\def\be{\begin{equation}}
\def\ee{\end{equation}}
\def\ben{\begin{eqnarray}}
\def\een{\end{eqnarray}}
\hsize=16.5truecm
\vsize=24truecm

\hoffset=-.3in

\thispagestyle{empty}
\begin{flushright} \ Oct. \ 1993 \\
KHTP-93-10/ SNUCTP 93-79\\
KATHEP-8-93
\end{flushright}
\begin{center}
 {\large\bf  Infinite Charge Algebra of  Gravitational Instantons
 }\\[.1in]
\vglue .5in
Jens Hoppe\footnote{E-mail address; be10@dkauni2.bitnet }
\vglue .5in
{\it Institut f\"{u}r Theoretiche Physik  \\
Universit\"{a}t Karlsruhe \\
P.O.Box 6980 \\
76137 Karlsruhe, Germany  }
\\[.2in]
{and}
\\[.2in]
Q-Han Park\footnote{ E-mail address; qpark@nms.kyunghee.ac.kr }
\vglue .5in
{\it Department of Physics, Kyunghee University\\
Seoul, 130-701, Korea}
\\[.2in]
{\bf ABSTRACT}\\[.2in]
\end{center}
\vglue .1in
Using a formalism of  minitwistors,
we derive infinitely many conserved charges for the $sl(\infty )$-Toda
equation which accounts for gravitational instantons with a rotational
Killing symmetry. These charges are shown to
form an infinite dimensional algebra through the Poisson bracket which is
isomorphic to two dimensional area preserving diffeomorphism with central
extentions.

\newpage

It has been known for some time that certain large N limits of  two dimensional
field theories yield higher dimensional field theories (see e.g.[1]).
 In particular, a large N limit of the two dimensional
 $sl(N)$-Toda equation becomes a three dimensional equation
 for a scalar field $u(w,\bar{w},t) ^{[1,2]}$,
 \be
 \pp\pb u = - \pp_{t}^{2}e^{u}\ ; \ \pp = {\pp \over \pp w}, \pb =
 {\pp \over \pp \wt }
 \ee
 which is also  the self-dual Einstein equation with a rotational
 Killing symmetry and the metric:$^{[3]}$
\be
ds^2 = {1 \over u_{,t}} (4 e^{u}dwd\wt + dt^{2}) +u_{,t}(d\theta +i u_{,w}dw +
i u_{,\wt }d\wt )^2                          \ .
\ee
 Eq.(1) as a large N limit of the  $sl(N)$-Toda equation is
 expected to possess infinite symmetries. Indeed, the infinitesimal action of
 such symmetries  has been obtained previously  and these
 were shown to form an algebra of area preserving diffeomorphisms.$^{[4]}$
 However, generators of such symmetries, i.e. conserved charges of
 Eq.(1), are not known explicitly except for few cases such as spin 2
charge.$^{[5]}$ Moreover the associated charge algebra which is essential in
understanding the quantum aspect of Eq.(1) is presently unknown.
 Even though  these charges are expected to  arise from large
N limits of conserved charges of  the  $sl(N)$-Toda equation, unlike the Toda
equation case,  the large N limit procedure for conserved charges is more
involved. As explained in this letter,  they are  correctly described by
the language of  twistor theory.

In this letter, we first derive conserved charges explicitly through the
 first order differential equations which determine the
infinitesimal symmetries of Eq.(1). Then, by defining a minitwistor space
for Eq.(1), we  provide  a general closed form of spin-s conserved charges.
These charges are also shown to form a symmetry  algebra via a  Poisson
bracket which is isomorphic to the centrally extended area preserving
diffeomorphisms.

We  first recall that the infinitesimal spin-s symmetry of Eq.(1),
\be
\pp\pb \d^{(s)}u = - \pp^{2}_{t} (e^{u}\d^{(s)} u) \ ,
\ee
is given by  the following recursive equations;$^{[4]}$
\ben
\d ^{(s)}u &=& \pp_{t} A^{(s)}_{0} \nonumber  \\
\pp_{t}A^{(s)}_{r-1} &=& (\pb + r\pb u )A^{(s)}_{r} \ ; \ r = 1,2, \cdots s-1
\een
with $A^{(s)}_{s-1}=f(\wt )$ and $f$ an arbitrary anti-holomorphic function.
Defining $\pp_{t}q \equiv u \ ; \ p \equiv  \pb q$ and
\be
K_{r} \equiv {1 \over \pp_{t}}(\pb + rp_{t}) \ \ ;
 \ \ M_{r} \equiv (\pb  - rp_{t}){1 \over \pp_{t}} \ ,
\ee
 the spin-s symmetry is formally given by
\be
\d ^{(s)} q = K_{1}\cdots K_{s-1}f(\wt ) \ .
\ee
If we define the fundamental Poisson bracket by
\be
\{ F(p(w, \wt ,t)) \ , \ G(p(w, \wt^{'},t^{'})) \} =
\int d\bar{v} d\tau {\d F \over \d p(w, \bar{v} , \tau )}{\pp \over
\pp \bar{v}}{\d G \over \d p(w, \bar{v} , \tau )} ,
\ee
the conserved spin-s charge $Q^{(s)}$ which acts as a generator of the
infinitesimal
symmetry of Eq.(4) may be introduced as follows:
\ben
\d^{(s)}p &=& \pb [K_{1}\cdots K_{s}f(\wt )]
= \{ p \ , \ \int   d\vt  f(\vt )Q^{(s)}(\vt ) \}  \nonumber \\
               &=& \pb [\int  d\vt  f(\vt ) {\d Q^{(s)}(\vt ) \over \d p(w, \wt
,  t) }]  \ ,
\een
or
\ben
\d^{(s)}q = \int d\vt f(\vt ){\d Q^{(s)}(\vt ) \over \d p(w,\wt ,t) } &=&
K_{1}\cdots K_{s-1}f(\wt )  \nonumber \\
&=& \int d\vt d\tau {\d p(w, \vt , \tau ) \over \d p(w, \wt ,  t ) }K_{1}
\cdots K_{s-1}f(\vt ) \nonumber \\
&=& \int d\vt  d\tau f(\vt )M_{s-1}(w, \vt ,\tau )\cdots M_{1}{ \d p(w, \vt ,
\tau ) \over \d p(w, \wt ,  t )} \
{}.
\een
This  implicit expression for the spin-s charge can be solved explicitly
for lower spin cases. For example,
\ben
Q^{(2)} &=& \int d\tau  [{1 \over 2}p^2 -\tau \pb p   ] \nonumber \\
Q^{(3)} &=& \int d\tau [{1 \over 3}p^3 + p\pb  \pp_{\tau }^{-1}p - \tau \pb p
p + {\tau^2 \over 2}\pb^{2}p ] \nonumber \\
Q^{(4)} &=& \int d\tau [{1 \over 4}p^4 - \tau p^2 \pb p + p^2 \pb\pp_{\tau
}^{-1}p-
p\pb p \pp_{\tau }^{-1}p \nonumber \\
&+& {\tau^{2} \over 2}(p\pb^{2} p + (\pb p )^{2}) + \tau \pb^{2}p\pp_{\tau
}^{-1}p +(\pb \pp_{\tau }^{-1}p )^{2} -
{\tau^{3} \over 6}\pb^{3}p ]
\een
Though one could essentially solve Eq.(9) for an arbitrary spin-s charge, in
practice this is
quite difficult. Moreover, the computation of the algebra among different
charges
through the Poisson bracket is much more involved. Instead
of trying to solve Eq.(9) for an arbitrary spin, in the following,
 we give yet another description of $Q^{(s)}$; i.e. `the minitwistor method'
 which provides a natural explanation for the recursive relations in Eq.(4)
 as well as other properties of  Eq.(1). Such a description essentially
 arises from the identification of Eq.(1) as the Einstein equation with a
rotational Killing symmetry.

 In order to do so,  we recall that most general self-dual Einstein equation
can be
given by$^{[6]}$
\be
\O_{,y\yt }\O_{,z\zt } - \O_{,y\zt }\O_{,z\yt } = 1
\ee
which is the same as the integrability of the linear equations:
\be
[ \l \pp_{\yt } + \O_{,\yt z}\pp_{y} - \O_{,\yt y}\pp_{z} ]\Psi = 0 \ ; \
 [ \l \pp_{\zt } + \O_{,\zt z}\pp_{y} - \O_{,\zt y}\pp_{z} ]\Psi = 0
\ee
where subscripts with a comma denote partial differentiation and $\l  \in
P^{1}$ is a
constant.
We impose a rotational Killing symmetry on the metric by assuming
$\O = \O (z, \zt , r=y\yt )$ and  change variables,
\be
 e^u \equiv r \equiv y\yt ,  \ w \equiv \zt , \  \wt \equiv z , \
  \O = \O(r,z,\zt ), \  t \equiv r\O_{,r} , \  s \equiv y \ ,
\ee
as well as coordinates,
 $(y,z,\yt ,\zt ) \rightarrow (w, \wt ,t, s )$, so that
\ben
\pp_{y} &=& {1 \over su_{,t}}\pp_{t} + \pp_{s } \ ; \
\pp_{\yt } = {s \over r_{,t}}\pp_{t} \nonumber \\
\pp_{z} &=& \pp_{\wt } -{r_{,\wt }\over r_{,t}}\pp_{t} \ ; \
\pp_{\yt } = \pp_{w} - {r_{,w} \over r_{,t}}\pp_{t} \nonumber \\
\pp_{r} &=& {1 \over r_{,t}}\pp_{t} \ ; \ t_{,r} = {1 \over r_{,t}} \ ,
\een
 Eq.(11) reduces  to Eq.(1) after some calculation and Eq.(12) becomes
 \ben
 ( \pp_{\wt } - {1 \over \h }\pp_{t} -  \pp_{\wt }u\h\pp_{\h }
) \Psi &=& 0
\nonumber \\
( \partial_{w } + e^{u} \h \partial_{t}  - e^{u}_{,t}\h^{2}\partial_{\h }
) \Psi &=& 0 \ ; \ \h = {1\over \l s } \ .
\een
Note that $\Psi = \Psi(w,\wt ,t,\h =  1/ \l s)$ depends on $\h $ while $u$
does not.
 In fact, Eq.(15) is a defining
equation of the twistor space for the $sl(\infty )$-Toda equation which is
known
as the minitwistor space. To be more specific, let $F $ and $ S$ be solutions
of Eq.(15),i.e.
\be
( \pp_{\wt } - {1 \over \h }\pp_{t} -  \pp_{\wt }u\h\pp_{\h }
) F = 0
= ( \partial_{w } + e^{u} \h \partial_{t}  - e^{u}_{,t}\h^{2}\partial_{\h }
) F
\ee
and the similar equations for $S$ with the asymptotic boundary conditions
\be
F(\h \rightarrow 0 ) \rightarrow {1 \over \h } \ ; \ S(\h \rightarrow 0 )
 \rightarrow \wt                 \ .
\ee
Then, $(F, S)  $ provide local coordinates of the minitwistor space
corresponding to rotationally symmetric gravitational instantons. In the case
$u=0$, which describes a three dimensional flat space $R^{2+1}$  degeneratedly
embedded into $R^{2+2}$, $F$ and $S$ become
\be
F = {1 \over \h } \ \ ; \ \ S = \wt + t\h - w\h^2 \ .
\ee
These are precisely the local coordinates of the minitwistor space $TP^{1}$, a
 tangent bundle over  $P^1 $, which corresponds to $R^{2+1}$.$^{[7]}$
Therefore,  $(F,S)$
generalize the minitwistor space to the curved metric case. This  also agrees
with the work by Jones and Tod$^{[8]}$ where the minitwistor space for
Einstein-Weyl spaces
was identified with the factor space of a twistor space by a holomorphic vector
field.

Having identified Eq.(15) as a defining equation of the minitwistor space for
the
$sl(\infty )$-Toda equation,  we may write the most general solutions for
$A^{(s)}_{r} $ in Eq.(4)
in terms of    contour integrals  over the minitwistor space;
\be
A^{(s)}_{r} = {1 \over 2\pi i}\oint_{\G }d\h \h^{r-1}F^{s-1}f(S) \ ; \ r = 0,1,
\cdots , s-1
\ee
where the contour $\G $ encloses $\h = 0$ and $f(S)$ is an arbitrary function.
A straightforward
calculation shows that   $A^{(s)}_{s-1} = f(\wt )$ and $A^{(s)}_{r}$ indeed
satisfies Eq.(4).
In particular,
\be
\d^{(s)}u = \pp_{t}A^{(s)}_{0} =
 {1 \over 2\pi i }\pp_{t}\oint_{\G }{d\h \over \h }F^{s-1}f(S) \ .
\ee
 This, when combined with eq.(8), gives
\be
\int d\vt  f(\vt ){\d Q^{(s)} \over \d p(w, \wt , t) } =
{1 \over 2 \pi i }\oint_{\G }{d \h \over \h }F^{s-1}f(S)
\ee
which is a contour integral expression of Eq.(9).
In fact, Eq.(21) could be understood as a large N limit of conserved charges of
$sl(N)$-Toda equation where the contour integral replaces the trace.
On the other hand, Eq.(21) manifests a sheaf
cohomological nature of  conserved charges in such a way that the integrand of
 the contour integral represents an element of the sheaf cohomology group
$H^{1}(T, {\cal O}( - 2  ))$ of the minitwistor space $T.^{[1]}$

Finally, we consider the charge algebra through the Poisson bracket eq.(7).
It has been shown in [4] that the infinitesimal symmetries
 given in Eq.(4) close under commutation so as to form an algebra s.t.
\be
[  \ \d^{(k)}_{f} \ , \ \d^{(l)}_{g} \ ] q = \d^{(k)}_{f} \d^{(l)}_{g} q -
\d^{(l)}_{g}d^{(k)}_{f} q =
 \d^{(k+l-2)}_{h}q
\ee
where $h = (k-1)fg^{'} - (l-1)f^{'}g $. This can be proved directly by using
the recursive relation Eq.(4) or by using the minitwistor formalism and
the sheaf cohomology.$^{[1]}$ On the other hand,
since
\be
\d^{(k)}_{f} \d^{(l)}_{g} q = \d^{(k)}_{f} [ {\d \over \d p}\int g Q^{(l)} ] =
\{ {\d \over \d p}\int g Q^{(l)}  \ , \ \int f Q^{(k)} \} \ ,
\ee
Eq.(22) gives rise to
\be
[  \ \d^{(k)}_{f} \ , \ \d^{(l)}_{g} \ ] q = {\d \over \d p} \{ \int gQ^{(l)} \
, \ \int fQ^{(k)} \}
  = {\d \over \d p} \int gQ^{(l +k-2)} = \d^{(k+l-2)}_{h}q \ .
\ee
This determines the charge algebra up to a central term $c$,
\be
\{ \int g Q^{(l)} \ , \ \int fQ^{(k)} \} = \int hQ^{(l+k-2)} + c \
\ee
which is precisely the algebra of centrally extended area preserving
diffeomorphisms. In order to determine the central charge in  Eq.(25),
we use the fact that the cocycle terms of area preserving diffeomorphisms
over genus $g$  surface have $2g$ independent cocycle terms.$^{[9]}$
 For a cylinder,  $g = 1/2$ and therefore we have only one central charge
up to a multiplicative constant. We find that in equation (25),
\be
c =  \int d\tau {\tau^{k+l-2} \over (k-1)!(l-1)!}
\int d \bar{v} {\partial^{l-1 } g \over \partial \bar{v}^{l-1} }
{\partial^{ k }f \over \partial \bar{v}^{k}} \
\ee
which agrees with the result from the lower spin calculation.

Let us conclude with a remark concerning the possibility$^{[10]}$ to transform
Eq.(1) into a `Monge-Amp\`{e}re form' (and containing the exponential of one of
the
independent, rather than the dependent, variable),
\be
(V_{22}V_{33} -V_{23}^{2}) + (V_{11}V_{33} - V_{13}^{2}) + (V_{11}V_{22}
-V_{12}^{2})
e^{p_{3}} = 0
\ee
with $V_{ij} := {\pp^{2} V \over \pp p_{i}\pp p_{j}} $, and $V = V(p_{1},
p_{2}, p_{3})$ being
related to $u(w = {x_{1} + ix_{2} \over 2}, \bar{w} = {x_{1} - ix_{2} \over 2}
, t = x_{3} ) = {\pp \over
\pp x_{3} }q(x_{1}, x_{2}, x_{3}) $ by a Legendre transformation, i.e. via
inverting $p_{i}(x) =
{\pp  \over  \pp x_{i}} q(x_{1}, x_{2}, x_{3})$ to obtain $x_{i}(p) = {\pp
\over \pp p_{i}}V(p_{1},p_{2},p_{3})$.
  \vglue 1.in
{\bf 	ACKNOWLEDGEMENT }
\vglue .2in
We would like to thank the Deutsche Forschungsgemeinschaft (J.H), the Korean
Science and Engineering Foundation (J.H. and Q.P.)   and the Program of Basic
Science Research, Ministry of Education(Q.P.),  for financial support.
\vglue .3in

\def\Item{\par\hang\textindent}

{\bf REFERENCES }
\vglue .1in
\Item {[1]} Q-H.Park, Phys.Lett.{\bf 238B} 287, (1990); Phys.Lett.{\bf257B},
(1991) 105 ;  Int.J.Mod.Phys.A, Vol. 7,No. 7 (1992) 1415 ;  I.Bakas, `Area -
Preserving
Diffeomorphisms and Higher Spin Fields in Two Dimensions', in `Supermembranes
and  Physics in 2+1 Dimensions', World Scientific 1990 (eds. M.Duff,C.N.Pope
and
E.Sezgin); K.Takasaki and T.Takebe,Lett.in Math.Phys.
{\bf 23} (1991) 205;  J.Hoppe, `Lectures on Integrable Systems' Springer 1992;
 J.Hoppe,M.Olshanetsky and S.Theisen,Commun.Math.Phys.{\bf 155} (1993) 429;
K.Takasaki,`Nonabelian KP Hierarchy with Moyal Algebraic Coefficients,
KUCP-0062/93 .
\Item {[2]} O.I.Bogoyavlenskii, Math.USSR Izv.{\bf 32} (1989) 245; M.V.Saveliev
and A.M.Vershik, Commun.Math.Phys.{\bf 126} (1989) 367 .
\Item {[3]} C.P.Boyer and J.D.Finley lll, J.Math.Phys.{\bf 23} (1982) 1126;
J.Gegenberg and A.Das, Gen.Relativ.Gravit.{\bf 16} (1984) 817 .
\Item {[4]} Q-H. Park, Phys.Lett.{\bf 236B} (1990) 429 .
\Item {[5]}  J.D.Finley lll and M.V.Saveliev, Phys.Lett.{\bf 162A} (1992) 1 .
\Item {[6]} J.F.Pleba\'{n}ski, J.Math.Phys.{\bf 16} (1975) 2396 .
\Item {[7]} N.J.Hitchin, Commun.Math.Phys. {\bf 83} (1982) 579; R.S.Ward,
J.Math.Phys.{\bf 30} (1989) 2246 .
\Item {[8]} P.E.Jones and K.P.Tod, Class. Quantum Grav. {\bf 2} (1985) 565 .
\Item {[9]} I.Bars,C.N.Pope and E.Sezgin, Phys.Lett.{\bf 210B} (1988) 85 .
\Item {[10]} Noted in collaboration with M.Bordemann .

\end{document}